\documentclass[10pt]{elsarticle}
\usepackage{epsfig}

\begin{document}

\title{Emergent geometry experienced by fermions in graphene in the presence of dislocations}

\author[AU,ITP]{G.E.~Volovik}

\author[UWO,ITEP]{M.A.~Zubkov
\footnote{Corresponding author, e-mail: zubkov@itep.ru}
 }

\address[AU]{Low Temperature Laboratory, School of Science and
Technology, Aalto University,  P.O. Box 15100, FI-00076 AALTO, Finland}

\address[ITP]{L. D. Landau Institute for Theoretical Physics,
Kosygina 2, 119334 Moscow, Russia}

\address[UWO]{The University of Western Ontario, Department of Applied
Mathematics,  1151 Richmond St. N., London (ON), Canada N6A 5B7}

\address[ITEP]{ITEP, B.Cheremushkinskaya 25, Moscow, 117259, Russia
}

\begin{abstract}
In graphene in the presence of strain the elasticity theory metric naturally appears. However, this is not the one experienced by fermionic quasiparticles. Fermions propagate in curved space, whose metric is defined by expansion of the effective Hamiltonian near the topologically protected Fermi point.  We discuss relation between both types of metric for different parametrizations of graphene surface. Next, we extend our consideration to the case, when the dislocations are present. We consider the situation, when the deformation is described by elasticity theory and calculate both torsion and emergent magnetic field carried by the dislocation. The dislocation carries singular torsion in addition to the quantized flux of emergent magnetic field. Both may be observed in the scattering of quasiparticles on the dislocation. Emergent magnetic field flux manifests itself in the Aharonov - Bohm effect while the torsion singularity results in Stodolsky effect.
\end{abstract}



\maketitle


\newcommand{\revision}[1]{{#1}}
\newcommand{\revisionB}[1]{{#1}}

\section{Introduction}


In this paper we discuss two different sources of effective metric:  elastic media
\cite{Bilby1956,Kroener1960,Dzyaloshinskii1980,KleinertZaanen2004,Vozmediano2010,Zaanen2010}
and topological matter  with Weyl or Dirac fermions \cite{Froggatt1991,Volovik2003,Horava2005}. We will consider the two dimensional graphene in the presence of elastic deformations as an example.

Let us start from the description of the deformations in $3D$ elastic media that are described by metric field $g_{ik}$:
\begin{equation}
g_{ik}= \frac{\partial X^l}{\partial x^i}
\frac{\partial X^l}{\partial x^k} = q^a_iq^a_k\,,
 \label{elasticity}
\end{equation}
where  $X^i({\bf x})=u^i({\bf x}) + x^i$ and $u^i({\bf x})$ is the displacement field, which describes the displacement of atoms from their equilibrium positions. Here we have introduced the (inverse) vielbein of elasticity theory $q_i^a$.   (in Ref.  \cite{Dzyaloshinskii1980} instead of the of the displacement of the lattice knots, the
system of the crystallographic coordinate planes $X^a({\bf x})$ has been used. In that case the vielbein
represents vectors orthogonal to the planes, i.e. the vectors of the reciprocal Bravais lattice.)
In terms of  the displacement field  $u^i({\bf x})$ one has:
\begin{equation}
q^a_k = \delta^a_k +\frac{\partial u^a}{\partial x^k}~~,~~g_{ik} = \delta_{ik} +2 \epsilon_{ik}
~~,~~ \epsilon_{ik}=\frac{1}{2}(\partial_i u^k + \partial_k u^i+ \partial_i u^l \partial_k u^l)
\,.
 \label{elasticity2}
\end{equation}
For small deformations, $\partial_i u^k\ll 1$, the displacement field is expressed in terms of the elastic deformations and rotations:
\begin{eqnarray}
&&\partial_j u_i(x) = \epsilon_{ij}(x) + \omega_{ij}(x)
 \label{elasticity3}
\\
&& \epsilon_{ij}= \frac{1}{2}(\partial_ju_i + \partial_i u_j ) , \quad    \omega_{ij}= \frac{1}{2}(\partial_ju_i - \partial_i u_j )
 \label{elasticity4}
\end{eqnarray}
 The curvature and torsion fields appear in the presence of  dislocations and disclinations
\cite{Bilby1956,Kroener1960,Dzyaloshinskii1980}.


The description of deformations in $2D$ media (graphene, the boundaries of topological insulators, etc) coincides with that of the $3D$ in case when there are no off - plane displacements of atoms (except that the indices take the values $i,j=1,2$ instead of $i,j = 1,2,3$). However, in the presence of the off plane displacements, when the considered surface is really curved, the description is different. Let us denote the three - vectors that give the coordinates of the surface points by
$X^K({\bf x})=u^K({\bf x}) - x^K$, where $K = 1,2,3$ and $x^3 = 0$ while $u^K({\bf x})$ is the displacement field. Then, metric of elasticity theory is given by
\begin{equation}
g_{ik}= \frac{\partial X^K}{\partial x^i}
\frac{\partial X^K}{\partial x^k}\,, \quad K = 1,2,3, \quad i,k = 1,2
 \label{elasticity}
\end{equation}
In terms of  the displacement field  $u^K({\bf x})$ one has:
\begin{equation}
g_{ik} = \delta_{ik} +2 \epsilon_{ik} = q_i^a q_j^a
~~,~~ \epsilon_{ik}=\frac{1}{2}(\partial_i u^k + \partial_k u^i+ \partial_i u^K \partial_k u^K)
\,.
 \label{elasticity2}
\end{equation}
Unlike the $3D$ case the inverse vielbein $q^a_k$ cannot be represented as $\frac{\partial X^a}{\partial x^k}$ because $X^K$ has three components.




The point nodes (zeroes) in the energy spectrum of fermionic excitations are also described by the vielbein, but of different origin. The point node protected by topology  is robust to the deformations of the system:
it does not disappear under deformation, though the Dirac cone is deformed and the position of the node changes. Topological theorem -- the Atiyah-Bott-Shapiro construction  \cite{Horava2005} -- dictates the linear expansion near the nodal point
 \begin{equation}
H=  - i \sigma^3 \sigma^a f^i_a ({\bf k}_i - {\bf K}_i) = {\bf i}^{ab}\sigma^b f^i_a ({\bf k}_i - {\bf K}_i), \quad {\bf i}^{12} = - {\bf i}^{21} = 1, \, {\bf i}^{11}={\bf i}^{22}=0
 \label{HUniform}
\end{equation}
Here ${\bf K}$ is the position of the Weyl or Dirac point, which for brevity will be further called  Dirac point. We use the representation for the coefficients ${\bf i}^{ab}f^i_a $  in order to match the particular expression for the Hamiltonian in graphene (see below).

In the inhomogeneous system the expansion parameters become  fields. The floating Dirac point
${\bf K}({\bf x})$ plays the role of the effective $U(1)$ gauge field, and the coordinate dependent matrix $f^i_a({\bf x})$ coincides with the effective vielbein $e^i_a({\bf x})$ up to the factor $e({\bf x}) = {\rm det}^{1/2}\Big( f^i_a({\bf x}) \Big)$ that defines invariant integration measure over coordinates (for the case of graphene see below, Sect. \ref{Secthom}):
\begin{equation}
f^i_a({\bf x}) = e({\bf x}) \, e^i_a({\bf x})\label{fe}
\end{equation}
Further in $2+1$ D case we shall refer to $f^i_a$ as to the zweibein and to $e^i_a$ as to the vielbein. The hamiltonian has the form:
 \begin{equation}
H=\frac{1}{2} {\bf i}^{ab}\sigma^b \left[f^i_a({\bf x})  \left(-i\partial_i - {\bf K}_i({\bf x})\right)+
 \left(-i\partial_i - {\bf K}_i({\bf x})\right) f^i_a({\bf x})\right] \,.
 \label{Heffective}
\end{equation}
So, close to the nodes the fermionic excitations behave as Weyl or Dirac  particles moving in the presence of the emergent vielbein and emergent gauge field. In the non - homogenious case in addition to these two fields the emergent spin connection may appear, so that we deal with the emergent Riemann - Cartan geometry. However, in case of graphene in the leading approximation it will not appear, so that we are left with the effective gravity described by the emergent vielbein with vanishing spin connection.

In the solid state materials with Dirac or Weyl points in the fermionic spectrum there is an interplay of the two metric fields: elastic and fermionic. In particular, the deformations or local rotations of the crystal,  which is described by the elastic gravity, causes the deformation of the fermionic vielbein, i.e. the fermionic gravity.
 We consider this effect on example of the 2D graphene.

\section{From bosonic to fermionic metric}

\subsection{Homogeneous strain. Dependence of results on the choice of coordinate system.}

Now let us come to the description of graphene.
We start with homogeneous elastic deformation $\partial_i u^k = const$  \cite{Oliva2013}.  The
 uniformly strained crystal is periodic,  and thus the expansion in Eq.(\ref{elasticity2}) with constant expansion parameters
remains valid. Since the pseudo-momentum $k_i$ remains a good quantum number, the homogeneous spectrum (\ref{HUniform}) is also valid.

There are two Fermi points in unstrained graphene $\pm {\bf K}^{(0)}$. In the presence of elastic deformations the positions of these two points are changed and are given by  ${\bf K}^{\pm} = \pm {\bf K}$.  The definition of ${\bf K}$ depends on the choice of coordinate system, i.e. on the parametrization of graphene surface. In \cite{VolovikZubkov2014} the parametrization typical for the elasticity theory is used. Namely, in the chosen parametrization of the surface the coordinates of Carbon atoms are the same as in the unperturbed honeycomb lattice. In this reference frame we have ${\bf K}   \approx   {\bf K}^{(0)} + {\bf A}$, where $\bf A$ is to be interpreted as the emergent $U(1)$ gauge field:
\begin{eqnarray}
A_y&=&-\frac{\beta}{2a}(\epsilon_{xx} - \epsilon_{yy}),~~  A_x=-\frac{\beta}{a}\epsilon_{xy}  \,,
 \label{TetradGaugeEffectiveA0}
\end{eqnarray}
where $\beta$ is the material Gruneisen parameter while $a$ is the lattice spacing.
The given parametrization is natural for the curved graphene surface. We refer to it as to the accompanying reference frame.

However, if the surface $\Sigma$ is curved only slightly and remains close to the ideal plane $\Sigma_0$, the other parametrization is preferred, which is called typically the "laboratory reference frame". In this parametrization the coordinates of Carbon atoms are given by their projections to the plane $\Sigma_0$. Let us choose the third coordinate axis orthogonal to $\Sigma_0$. Then the transformation between the two parametrizations of $\Sigma$ is given by:
\begin{equation}
X^k({\bf x})=u^k({\bf x}) + x^k, \quad k = 1,2
\end{equation}
Here the laboratory reference frame coordinates are denoted by $X^k$ while the coordinates of the original reference frame (accepted in \cite{VolovikZubkov2014}) are denoted by $x^k$.

\subsection{Homegenious case. Expressions for emergent vielbein and emergent gauge field.}
\label{Secthom}

In laboratory reference frame the value of ${\bf K}$  has two contributions:
\begin{eqnarray}
{\bf K}_i & \approx & \frac{\partial x^k}{\partial X^i} \Big({\bf K}_k^{(0)}+  {\bf A}_k\Big) \approx \Big(\delta_i^k -  \partial_i u^k \Big) \Big({\bf K}_k^{(0)}+  {\bf A}_k\Big)\nonumber\\ & \approx &  ({\bf K}_i^{(0)}-  \nabla_i({\bf u}\cdot {\bf K}^{(0)})) +    {\bf A}_i
 \label{DiracPosition1}
\end{eqnarray}
The first term in the right hand side of this expression contains the geometric contribution, which comes from the coordinate transformation of the original position ${\bf K}^{(0)}$ of the Dirac node in the non-deformed lattice. The second term depends on the material parameter $\beta$:
\begin{equation}
   A_y=-\frac{\beta}{2a}(\epsilon_{xx} - \epsilon_{yy})~~,~~  A_x=-\frac{\beta}{a}\epsilon_{xy}   \,,
 \label{A}
\end{equation}
where $a$ is the interatomic space.

The Hamiltonians for the particles near the two valleys  (correspond to the values of momenta close to ${\bf K}^{\pm}$) have the form \cite{VolovikZubkov2014}:
 \begin{eqnarray}
H_{-}&=&-\frac{1}{2} i \sigma^3 \sigma^a \left[f^i_a({\bf x})  \left(-i\partial_i + {\bf K}_i({\bf x})\right)+
 \left(-i\partial_i + {\bf K}_i({\bf x})\right) f^i_a({\bf x})\right]\nonumber\\
 H_{+}&=&-\frac{1}{2} i \sigma^2\sigma^3 \sigma^a \left[f^i_a({\bf x})  \left(-i\partial_i - {\bf K}_i({\bf x})\right)+
 \left(-i\partial_i - {\bf K}_i({\bf x})\right) f^i_a({\bf x})\right]\sigma^2
 \label{Heffective0}
\end{eqnarray}
The low energy effective action has the form:
\begin{equation}
S = \sum_{\pm}\int d^2x dt {\psi}^+_{\pm} \Big(i \partial_t - H_{\pm}\Big)\psi_{\pm},
\end{equation}
where $\psi_{\pm}, {\psi}^+_{\pm}$ are the independent fermion Grassmann variables that describe quasi - particles living near the Fermi points ${\bf K}^{\pm}$. Let us introduce the new variables $\Psi_{\pm}, \bar{\Psi}_{\pm}$:
\begin{eqnarray}
\Psi_- &=& \psi_-, \quad \bar{\Psi}_- = \psi^+_- \sigma^3\nonumber\\
\Psi_+ &=& \sigma^2 \psi_+, \quad \bar{\Psi}_- = \psi^+_+ \sigma^2\sigma^3
\end{eqnarray}
and the $2+1$ D (Minkowski) gamma - matrices:
\begin{equation}
\gamma^0 = \sigma^3, \quad \gamma^1 = -i\sigma^1, \quad \gamma^2 = -i\sigma^2
\end{equation}
These matrices satisfy
\begin{equation}
\{\gamma^{a}, \gamma^b\} = 2 \eta^{ab} = 2\,{\rm diag} \,(1,-1,-1), \quad a,b = 0,1,2
\end{equation}
In terms of these new variables the effective action has the form:
\begin{equation}
S =\frac{1}{2} \sum_{\pm} \, \int d^2x dt \, e\,\bar{\Psi}_{\pm} e_a^{\mu}\gamma^a \Big(i \partial_{\mu} \pm K_{\mu}\Big)\Psi_{\pm} + (h.c.), \quad K_{\mu} = (0,{\bf K})
\end{equation}
Here $(h.c.)$ means hermitian conjugated expression. According to our definition the hermitian conjugation applied to $\Psi$ gives $\bar{\Psi}$. The effective $2+1$ D  vielbein $e^{\mu}_a$ is related to the zweibein $f^i_a$ as follows:
\begin{eqnarray}
 e_a^{\mu} & = & f_a^{\mu} \, {\rm det}^{-1/2} f^i_a, \quad a,\mu = 1,2, \quad e_0^{\mu}  =  \delta_0^\mu\, {\rm det}^{-1/2} f^i_a \nonumber\\
 e^a_{\mu} & = & \Big[e^{-1}\Big]^a_{\mu}, \quad a,\mu = 1,2,  \quad e_\mu^{0}  = \delta^0_{\mu}\,{\rm det}^{1/2} f^i_a\nonumber\\
 e & = & {\rm det} \, e^a_{\mu} = {\rm det}^{1/2} f^i_a
\end{eqnarray}

We may relate parameters $f^i_a $ of the fermionic spectrum with the elastic deformation $u^i$.
In  the non-disturbed  graphene  the  fermionic spectrum near the Dirac point is isotropic,  being determined by the Fermi velocity $v_F$ of quasiparticles.
As well as for ${\bf K}$ the definition of $f^i_k$ depends on the choice of coordinate system, i.e. on the parametrization of graphene surface. In the accompanying reference frame we have (Eq. (3.4) of \cite{VolovikZubkov2014}):
\begin{equation}
 f^i_a \approx v_F\left(\delta^i_a-\beta \epsilon_{ia} \right)\,.
 \label{Connection0}
\end{equation}
The second term in the right hand side of Eq.(\ref{Connection0})
comes from the dependence of the hopping elements on the distances between the atoms in the lattice, which are changed under the deformation of the lattice. The coefficient
$\beta$ is the non-geometric material parameter, which is called the Gr\"uneisen parameter.

Applying the coordinate transformation to Eq. (\ref{Connection0}) we come to the following expression for $f^i_a$ in the laboratory reference frame:
\begin{equation}
 f^i_a ={\rm det}^{-1} \Big(\frac{\partial X}{\partial x}\Big) v_F\left(\delta^k_a-\beta \epsilon_{ka} \right)\,\frac{\partial X^i}{\partial x^k}\approx v_F\left(\delta^i_a(1-\partial_k u_{k}) + \partial_i u_a -\beta \epsilon_{ia} \right)
 \label{Connection1}
\end{equation}
If the term $\partial_i u^K \partial_j u^K$ may be neglected compared to $\partial_j u^a$, then we come to
\begin{equation}
 f^i_a \approx v_F\left(\delta^i_a(1-\epsilon_{kk}) +  \epsilon_{ia} + \omega_{ia}-\beta \epsilon_{ia} \right),  \quad    \omega_{ij}= \frac{1}{2}(\partial_ju_i - \partial_i u_j )
 \label{Connection11}
\end{equation}
In the particular case, when the off - plane displacements of graphene atoms are absent at all, this result was independently obtained in \cite{Oliva2013,Oliva2014}.
The terms $\delta^i_a(1-\partial_k u_{k}) + \partial_i u_a$ in the rhs of Eq.(\ref{Connection1}) are of the geometric origin: they come from the coordinate transformation from the original non-disturbed to the strained graphene.

\subsection{Inhomogeneous strain}

The extension to the non-uniform deformations is achieved by localization of the expansion parameters according to Eq.(\ref{Heffective}), which means that the distorsion $\partial_i u_j$ should be considered as coordinate dependent. Since $\nabla\times  ({\bf K}^0-  \nabla({\bf u}\cdot {\bf K}^0))=0$, the first term in the rhs of Eq.(\ref{DiracPosition1}) does not produce the pseudo-magnetic field. It can be eliminated by
gauge transformation, which leaves only the gauge field ${\bf A}({\bf x})$.  As a result one obtains the effective Hamiltonians for the fermions living near the two valleys of the form of
\begin{eqnarray}
H_{-}({\bf A})&=&-\frac{1}{2} i \sigma^3 \sigma^a \left[f^i_a({\bf x})  \left(-i\partial_i + {\bf A}_i({\bf x})\right)+
 \left(-i\partial_i + {\bf A}_i({\bf x})\right) f^i_a({\bf x})\right], \quad  H_{+}({\bf A})=\sigma^2 H_{-}(-{\bf A})\sigma^2 \label{Heffective2}
\end{eqnarray}
where the effective zweibein $f^i_a({\bf x})$ and the emergent gauge field ${\bf A}({\bf x})$
in graphene are expressed in "laboratory reference frame" as follows
\begin{eqnarray}
 f^i_a({\bf x}) & \approx & v_F\left(\delta^i_a(1-\partial_k u_{k}({\bf x})) + \partial_i u_a({\bf x}) -\beta \epsilon_{ia}({\bf x}) \right)
 \label{FermiTetrad}
\\
  A_y({\bf x})& =&-\frac{\beta}{2a}(\epsilon_{xx}({\bf x}) - \epsilon_{yy}({\bf x})),~~  A_x({\bf x})=-\frac{\beta}{a}\epsilon_{xy}({\bf x})  \,.
 \label{TetradGaugeEffective}
\end{eqnarray}
Transformation with $\partial_i u_j({\bf x}) = - \partial_j u_i({\bf x}) = \omega_{ij}({\bf x})$  describes rotations in the plane of graphene.
The pure rotations that do not depend on coordinates ($\omega_{ij} = {\rm const}$)  leave the zweibein isotropic. To see that we may in addition rotate the internal space accordingly thus giving rise to the transformation of the Pauli matrices $\sigma^a$.
As a result we come back to the isotropic Fermi velocity: $f^i_j({\bf r})  \approx v_F\left(\delta_{ik} + \omega_{ik} \right) \left(\delta_{kj} - \omega_{kj} \right) = v_F \delta_{ij}$.

In the accompanying reference frame (in which the Carbon atoms have the same coordinates as in regular honeycomb lattice) we have:
\begin{eqnarray}
 f^i_a({\bf x}) &\approx & v_F\left(\delta_{ia} -\beta \epsilon_{ia}({\bf x}) \right)\,,
 \label{FermiTetrad}
\\
  A_y({\bf x})&=&-\frac{\beta}{2a}(\epsilon_{xx}({\bf x}) - \epsilon_{yy}({\bf x})),~~  A_x({\bf x})=-\frac{\beta}{a}\epsilon_{xy}({\bf x})  \,.
 \label{TetradGaugeEffectiveA}
\end{eqnarray}

From Eq. (\ref{Heffective2}) it follows that the fermions living near to the two Fermi points ${\bf K}^+$ and ${\bf K}^-$ have opposite charges with respect to the emergent gauge field $\bf A$. This is in contrast to the real electromagnetic field: fermions at both valleys have the same electric charge.

\section{Emergent geometry in the presence of dislocations}

\subsection{Local map in the vicinity of the origin of the dislocation}

\subsubsection{Accompanying reference frame}

Let us consider how the emergent geometry is affected by dislocation (see also \cite{Guinea2014}). Both effective geometry of elasticity theory and the effective geometry experienced by fermions acquire the singular contributions to torsion concentrated at the origin $x^0$ of the dislocation. However, we shall see, that these contributions  are different.
At the origin $x^0$ of the dislocation the pair hexagon - hexagon may be substituted, for example, by the pair heptagon - pentagon. Then the extra sequence of hexagons is added along the line $\cal J$ that starts at the pentagon. For the low energy effective theory this results in cutting of the graphene surface along the line $\cal J$ that begins at $x^0$ and goes to infinity. Then the strip of a finite width is added along $\cal J$.  The resulting surface is sewn along the cut. As a result in the accompanying reference frame (where the coordinates of the atoms are the same as in the unperturbed honeycomb lattice) there is the uncertainty in the definition of the parametrization at the cut. This uncertainty gives nonzero value to the  following integral along the contour $\cal C$ surrounding $x^0$:
\begin{equation}
b^i = \int_{\cal C} d x^i
\end{equation}

Let us introduce vectors that connect a
vertex of the unperturbed honeycomb lattice with its neighbors:
\begin{equation}
{\bf l}_1= (-a,0),\qquad {\bf l}_2 =  (a/2,a\sqrt{3}/2),\qquad {\bf l}_3=
(a/2,-a\sqrt{3}/2)\label{uuu3}
\end{equation}
Also we
define the following vectors:
\begin{eqnarray}
&&{\bf m}_1= -{\bf l}_1 + {\bf l}_2, \qquad {\bf m}_3 = -{\bf
l}_3 + {\bf l}_1,  \qquad {\bf m}_2=
-{\bf l}_2 + {\bf l}_3
  \label{uuu_2}
\end{eqnarray}
One can check that the allowed values of Burgers vector are
\begin{equation}
{\bf b} = \sum_{i=1,2,3} N_i {\bf m}_i \label{burgers}
\end{equation}
with integer values $N_i$.
In the mentioned particular case when the pair hexagon - hexagon is substituted by the pair heptagon - pentagon the Burgers vector $\bf b$ is equal to ${\bf m}_k$ for $k = 1$, $2$, or $3$.

\subsubsection{Laboratory reference frame}

In laboratory reference frame the dislocation is described in a different way. This parametrization of graphene surface does not contain ambiguity. The ambiguity in the parametrization of accompanying reference frame appears through the displacement field defined as a function of $X^k$:
\begin{equation}
x^k = X^k - u^k(X)
\end{equation}
(Off - plane displacements do not enter.) The parametrization in accompanying reference frame is defined modulo the step - like discontinuity at the cut. Therefore, although the displacement field $u^i(X)$ has a step - like discontinuity concentrated along the cut, its derivative $\partial_j u^k$ is continuous. The Burgers vector is given by:
\begin{equation}
b^i = \int_{\cal C} d x^i = - \int_{\cal C} d u^i
\end{equation}
We may choose, for example, the following representation:
\begin{equation}
u^a = -  \phi \frac{b^a}{2\pi}  + u^a_{\rm cont},\label{using}
\end{equation}
where $\phi$ is the polar angle ($X_1= X^0_1 + r \, {\rm cos} \, \phi$, $X_2=X^0_2+ r \, {\rm sin} \, \phi$) while $u^a_{\rm cont }$ is continuous along the cut (i.e. it does not contain the discontinuity along the cut). However, $u_{\rm cont}$ may be undefined at the origin of the dislocation.

In the following we \revision{shall imply that the graphene surface is nearly flat} and shall consider expressions for torsion and magnetic field carried by the dislocation in the laboratory reference frame, where there is no ambiguity in parametrization.
It is useful to describe dislocation in this reference frame using the field of elastic vielbein that is regular along the cut and is undefined at the origin of the dislocation only. For small out of plane displacements we may identify $\delta_i^a+\partial_i u^a$  with the  (inverse) vielbein of elasticity theory similar to Eq. (\ref{elasticity2}). It is continuous out of the origin of the dislocation because $u^a$ is defined modulo $b^a$ at the cut. When the off - plane displacement cannot be neglected, tensor $q_i^a = \delta_i^a+\partial_i u^a$ cannot be identified with the elastic vlelbein, but it still can be used for the calculation of torsion and emergent magnetic field (see below). The Burgers vector is expressed through $q_i^a$ as
\begin{equation}
b^a =  - \int_{\cal C} q_i^a d X^i
\end{equation}
This means that
\begin{equation}
\partial_{1} q_{2}^a - \partial_{2} q_{1}^a = - b^a \delta^{(2)} (X-X^0)
\end{equation}

\revision{The first term in Eq. (\ref{using}) is specific for the dislocation. At the same time $u_{\rm cont}$ may be or may not be related to the dislocation depending on external conditions. For brevity we shall consider the  relatively simple case of suspended graphene in vacuum (when there are no external forces at all). Then the equilibrium values of $u^a_{\rm cont}$ are to be defined using elasticity theory (see below, Sect. \ref{sectelast}). There is the relatively simple solution of elasticity equations that corresponds to the vanishing off - plane displacements. It appears, that such a solution is singular at the origin of the dislocation. For the really curved graphene surface the elasticity equations are non - linear, and their solution is rather complicated. We do not consider it in the present paper.}

\revision{It is worth mentioning that in general case of the graphene layer placed on the substrate the connection between the substrate and the grapnene layer complicates the consideration. When the connection is sufficiently strong, expression for $u^a_{\rm cont}$ may not be predicted using simple elasticity equations. In this case $u^a_{\rm cont}$ should be considered as the given function of $X$, whose shape is fixed by external conditions. In principle, this function may also be singular at the position of the dislocation.}

\subsection{Elasticity equations}
\label{sectelast}

Elastic part of the free energy of the thin plate may be written as \cite{Landau, Katsbook}:
\begin{equation}
F = \frac{1}{2}\int d^2 X \Big(\kappa (\Delta h)^2 + \lambda \epsilon^2_{ll} + 2 \mu \epsilon_{ik} \epsilon_{ik} \Big), \label{F}
\end{equation}
where $\lambda$ and $\mu$ are the two - dimensional Lame coefficients, $h = u_3$, while $\kappa$ is bending rigidity \cite{Katsbook}.
We may neglect the term quadratic in $u_a$ in $\epsilon_{ij}$ but cannot in general case neglect the term quadratic in $h$. Therefore, the free energy is quadratic in $u_a$ ($a=1,2$) and contains up to the forth power of $h$. As a result the differential equations are linear in $u^a$ and non - linear in $h$.

\revision{As it was mentioned above, the elasticity equations may be solved rather easily in the case of  suspended graphene in vacuum (when all external forces including gravitational may be neglected). Then there is the simple solution that corresponds to vanishing off - plane displacements (see below). The variation of Eq. (\ref{F}) over $h$ gives equation that relates it with $\epsilon_{ij}$. The corresponding equation has the solution with $h = 0$.}

The variation over $u^a$ ($a=1,2$) gives \cite{Landau}:
\begin{equation}
\partial_k \epsilon_{ik} + \frac{\sigma}{1-\sigma} \partial_i \epsilon_{ll} = 0,\label{elaste}
\end{equation}
\revision{Here we introduced the two - dimensional parameter $\sigma$ as:
$\frac{\sigma}{1-\sigma}=\frac{\lambda}{\mu}$.
It is introduced in such a way, that Eq. (\ref{elaste}) has the form of the equation for the  $2D$ plate. Do not confuse, however, $\sigma$ with the notion of the $3D$ Poisson parameter that relates constriction coefficient in direction orthogonal to the plate plane with the constriction coefficient in the in - plane direction. In case of the one - atom layer of graphene this $3D$ coefficient does not have sense. It is also worth mentioning, that Eq. (\ref{elaste}) differs from the corresponding $3D$ equation, where instead of $\frac{\sigma}{1-\sigma}$ the factor $\frac{\sigma}{1-2\sigma}$ appears.}   Eq. (\ref{elaste}) together with the boundary conditions determines the values of the equilibrium in - plane displacement field $u^a$. In the presence of the dislocation we may substitute Eq. (\ref{using}) and obtain the
equation for the continuous part of $u^i$:
\begin{eqnarray}
0 &=& (\Delta \delta^{ik} + \frac{1+\sigma}{1-\sigma} \nabla^i \nabla^k)\Big(u^k_{\rm cont}-b^l {\bf i}^{kl} \frac{{\rm log}\,|X|}{2\pi}\Big)+ 2 b^l {\bf i}^{il} \delta^{(2)}(X),\nonumber\\ &&  \quad {\bf i}^{12} = - {\bf i}^{21}=1, \quad {\bf i}^{11}={\bf i}^{22} = 0, \quad \hat{X}^i = \frac{X^i}{|X|}, \label{elastU}
\end{eqnarray}
Here we place the origin of the dislocation at $X^i = 0$. We come to the following solution:
\begin{eqnarray}
 u^k_{\rm cont}(X) &=&  b^l {\bf i}^{kl} \frac{({\rm log} \frac{|X|}{2 R} +\gamma)}{2\pi}- \frac{1}{\Delta}( \delta^{ik} - \frac{1+\sigma}{2}\frac{ \nabla^i \nabla^k}{\Delta})   2 b^l {\bf i}^{il} \delta^{(2)}(X) \label{ucontsol0}  \\
&=& - \frac{b^l {\bf i}^{kl}}{2\pi}({\rm log} \frac{|X|}{2 R} +\gamma) +  (1+\sigma) \nabla^i \nabla^k  \frac{ b^l {\bf i}^{il}}{8\pi}\, ({\rm log} \frac{|X|}{2 R} +\gamma-1)|X|^2\nonumber
\end{eqnarray}
The inverse Laplace operator $\Delta^{-1}$ is to be defined taking into account boundary conditions and the finite size of graphene sample. For the sample of linear size $R$ we come to the following expressions:
\begin{equation}
- \Big[\Delta^{-1}\Big]_X = \int \frac{d^2 K}{(2\pi)^2}\frac{e^{i K^a X^a}}{|K|^2}\approx  - \frac{1}{2\pi}\, ({\rm log} \frac{|X|}{2 R} +\gamma), \quad |X| \ll R
\end{equation}
($\gamma$ is the Euler constant), and
\begin{equation}
\Big[\Delta^{-2}\Big]_X = \int \frac{d^2 K}{(2\pi)^2}\frac{e^{i K^a X^a}}{|K|^4}\approx \frac{R^2}{4\pi} +  \frac{1}{8\pi}\, ({\rm log} \frac{|X|}{2 R} +\gamma-1)|X|^2, \quad |X| \ll R
\end{equation}
Up to an irrelevant constant we have
\begin{eqnarray}
 u^k_{\rm cont}(X) &=&  -\frac{1-\sigma}{2}  b^l {\bf i}^{kl}   \frac{1}{2\pi}  {\rm log} \frac{|X|e^{\gamma}}{2 R}+  b^l {\bf i}^{il} (1+\sigma) \frac{\hat{X}^i \hat{X}^k}{4\pi} \label{ucontsol1}
\end{eqnarray}

This expression should be regularized both at $|X|\rightarrow 0$ and at $|X| \rightarrow \infty$. The considered effective field theory fails at the distances of the order of the lattice spacing $a$. Actually, for $|X| \le r$, where $1/r > 1/a$ is the ultraviolet cutoff, the value of $u_{\rm cont}^k$ is not given by Eqs. (\ref{elastU}), (\ref{ucontsol1}).   The effects of the finite size of the graphene sample are also strong. Notice, that while the values of $u_{\rm cont}^k$ may be large due to the large difference between the two scales $r$ and $R$, its derivatives are small for sufficiently small $b$ because after the differentiation the expression in Eq. (\ref{ucontsol1}) tends to zero at $|X|\rightarrow \infty$.

\subsection{Expression for emergent magnetic field carried by the dislocation}

According to Eq. (\ref{DiracPosition1}) in laboratory reference frame the emergent electromagnetic field has the form
\begin{eqnarray}
A_i & \approx & -  \nabla_i({\bf u}\cdot {\bf K}^{(0)}) +    {A}^{\rm acc}_i
 \label{DiracPosition3}
\end{eqnarray}
where ${\bf K}^{(0)}_i = \frac{4\pi}{3 \sqrt{3}a} \delta_{i2}$ is the position of the unperturbed Fermi point while ${A}^{\rm acc}_i$ is given by Eq. (\ref{TetradGaugeEffectiveA}).

In order to calculate the emergent magnetic field we use integral expression
\begin{equation}
\int_{\cal S} H dx^1 \wedge dx^2 \equiv \int_{\partial {\cal S}} A_k dX^k \label{AdefI}
\end{equation}
For the considered solution of elasticity equations with $\partial_k h \partial_l h = 0$, when $\partial_i u^a \partial_j u^a$ may be neglected, we represent the right hand side of this expression as follows
\begin{equation}
\int_{\partial {\cal S}} A_k dX^k =  b^i {\bf K}^{(0)}_i-\frac{\beta}{2 a}\int_{\partial {\cal S}} \Big(q^2_1 dX^1 + q^1_2 d X^1 + q^1_1 d X^2 -q^2_2 d X^2 \Big) \label{AdefI2}
\end{equation}
The first term in this expression gives the following singular contribution to magnetic field:
\begin{equation}
H_{{\rm sing}} \approx  b^i {\bf K}^{(0)}_i \delta^{(2)}(x-x^0),\label{Hsing}
\end{equation}
The unperturbed Fermi point is defined up to the transformation ${\bf K}^{(0)} \rightarrow {\bf K}^{(0)} + {\bf G}$, where ${\bf G}$ is the vector of inverse lattice. This corresponds to the change of the magnetic flux by
$\Delta \Phi={\bf b}\cdot{\bf Q} = 2\pi N$, where $N$ is integer. Such change of the magnetic flux is unobservable for the Dirac fermions.

The contribution coming from the second term of Eq. (\ref{using}) (given by Eq. (\ref{ucontsol1})) to Eq. (\ref{AdefI2}) vanishes for small $\cal S$. Therefore, the magnetic field originated from $u_{\rm cont}^k$ is regular and is given by
\begin{equation}
H_{{\rm cont}} = \frac{\beta}{2a}\Big(2 \partial_1\partial_2 u_{\rm cont}^2 + (\partial_2^2-\partial_1^2) u^1_{\rm cont}\Big), \label{Hcont}
\end{equation}
where $u_{\rm cont}^k$ is given by Eq (\ref{ucontsol1}). As it was mentioned above, both expressions of Eq. (\ref{ucontsol1}) and Eq. (\ref{Hcont}) are valid for $|X-X^0| > r$, where $1/r$ is the scale at which the field theoretical description starts to work. At the same time Eq. (\ref{Hsing}) originates from the integral over the closed contour taken along the path that may be placed far from the origin of the dislocation. Therefore, this singular term is not affected by the theory working at small distances $\le r$ and it gives the value of emergent magnetic flux carried by the dislocation. This magnetic flux appears to be proportional to $\frac{2 \pi}{3}$: for the value of Burgers vector given by Eq. (\ref{burgers}) it is given by (see also \cite{gauge_dislocation})
\begin{equation}
\Phi = \Big(N_1 + N_2 +  N_3\Big)\frac{2 \pi}{3}\label{magnflux}
\end{equation}
The observation of this phase corresponds to the ordinary Aharonov - Bohm effect and may be performed using scattering of a quasiparticle on the dislocation.

\subsection{Expression for torsion carried by the dislocation}

Dislocation produces the delta - functional contribution to torsion in space with metric of elasticity theory.
However, it is not the torsion experienced by fermionic quasiparticles. In the absence of dislocation the torsion tensor is defined as
\begin{eqnarray}
T^a_{jk} &\equiv & \partial_{[j}  e^a_{k]},
\end{eqnarray}
where $e^a_k$ is the (dimensionless) inverse $2+1$ D vielbein related to $f_a^i$ according to Eq. (\ref{fe}).
In the presence of dislocation we use integral representation:
\begin{equation}
\frac{1}{2}\int_{\cal S} T_{ij}^a dX^i \wedge dX^j \equiv \int_{\partial {\cal S}} e^a_k(X) dX^k \label{TedefI}
\end{equation}
In order to calculate torsion at $X^0$ we should choose $\cal S$ as its small vicinity.

In the accompanying reference frame for the components of $e^a_k$ with $k,a = 1,2$ we have:
\begin{eqnarray}
e^a_k(x) &=& \Big(\delta^k_a - \beta \epsilon_{ka}\Big)^{-1}{\rm det}^{1/2}\Big(\delta^k_a - \beta \epsilon_{ka}\Big)\nonumber\\ &\approx & (\delta^a_k + \beta \epsilon_{ka})\,(1-\frac{\beta}{2} \epsilon_{ii})\nonumber\\ &\approx & \Big(\delta^a_k + \beta \epsilon_{ka}  -  \frac{\beta}{2} \epsilon_{ii}\delta^a_k\Big)
\end{eqnarray}
The values of $e^a_k$ in laboratory reference frame are given by
\begin{eqnarray}
e^a_k(X) = \frac{\partial x^j}{\partial X^k} e^a_j(x) &\approx & \Big(\delta^a_k -  \partial_k u^a  + \beta \epsilon_{ka}  -  \frac{\beta}{2} \epsilon_{ii}\delta^a_k\Big)\label{elab}
\end{eqnarray}
In principle, we should also transform derivatives entering $ \epsilon_{ka}$: $\frac{\partial}{\partial x^k} \rightarrow \frac{\partial X^i}{\partial x^k}\frac{\partial}{\partial X^i}$. However, we are able to substitute here $\frac{\partial X^i}{\partial x^k}$ by $\delta^i_k$ in the approximation linear in the in - plane displacement $u^i, i = 1,2$.

\revision{According to Eq. (\ref{TedefI}) torsion is related to the circulation of the inverse vielbein $e^a_k$ along the closed contour.   Let us represent $e^a_k$ in terms of the inverse vielbein $q_k^a$ of elasticity theory for the case, when $\partial_k u^a \partial_l u^a$ ($a=1,2$) may be neglected (recall that $\partial_k h \partial_l h = 0$ for the considered solution):
\begin{eqnarray}
\int e^a_k(X)d X^k &\approx & \int \Big( - q^a_k(X) + \frac{\beta}{2}(q^a_k(x) + q^k_a(X) - q^j_j(X) \delta^a_k)\Big) dX^k \label{eq}
\end{eqnarray}
The circulation of $q^a_i$ gives the following singular contribution to torsion:
\begin{eqnarray}
T^a_{12, {\rm sing}} &\approx & (1-\beta/2) b^a \delta^{(2)}(X-X^0)\label{tsing}
\end{eqnarray}
The remaining part of Eq. (\ref{eq}) looks non - covariant. This is because Eq. (\ref{FermiTetrad}) was obtained in a particular fixed gauge. This remaining part gives the following contribution to $\int e^a_k d X^k$:
\begin{equation}
\frac{\beta}{2}\int \Big(q^k_a dX^k - q^j_j dX^a\Big)=\frac{\beta}{2} \int {\bf i}^{ad} \nabla_b u^d {\bf i}^{bc}d X^c =  \frac{\beta}{2} \int_0^{2\pi} {\bf i}^{ad}X^b \nabla_b u^d  d \phi \label{reg}
\end{equation}
The contribution of the first term of Eq. (\ref{using}) to this expression vanishes. Therefore, Eq. (\ref{reg}) gives the following contribution to torsion:
\begin{equation}
T^a_{12, {\rm cont}} = \frac{\beta}{2} {\bf i}^{fc} {\bf i}^{ad}{\bf i}^{bc}\nabla_f \nabla_b u^d  = \frac{\beta}{2} {\bf i}^{ad}\Delta u^d, \quad \Delta = \partial_1^2 + \partial_2^2
\end{equation}
In the case considered above in Sect. \ref{sectelast}, when the elasticity theory may be applied, and there is the solution with $h=0$, $u_{cont}^a$ is given by Eq. (\ref{ucontsol1}). The corresponding part $T_{12, {\rm cont}}^a(X)$ of torsion is given by  (for $X\ne 0$):
 \begin{eqnarray}
T_{12, {\rm reg}}^a(X) &=&  \frac{\beta(1+\sigma)}{2} \frac{\delta^{ik} - 2\hat{X}^i\hat{X}^k}{2\pi |X|^2}b^l {\bf i}^{il} {\bf i}^{ak}, \quad X\ne 0 \label{ucontsol4}
\end{eqnarray}
Here $u_{\rm cont}$ is the part of the displacement vector that does not contain discontinuity along the cut but that is undefined at the origin of the dislocation.
This expression is not well - defined at $X\rightarrow 0$. In order to calculate the part of torsion originated from $u_{\rm cont}$ localized on the dislocation we should substitute into Eq. (\ref{reg}) the expression for $u_{\rm cont}^k$ given by Eq. (\ref{ucontsol1}):
\begin{eqnarray}
 u^k_{\rm cont}(X) &=& -\frac{1-\sigma}{2}  b^l {\bf i}^{kl}   \frac{1}{2\pi}  {\rm log} \frac{|X|e^{\gamma}}{2 R}+  b^l {\bf i}^{il} (1+\sigma) \frac{\hat{X}^i \hat{X}^k}{4\pi} \label{ucontsol122}
\end{eqnarray}
One can check that the second term here does not contribute to Eq. (\ref{reg}.) The first term been substituted to Eq. (\ref{reg}) gives the contribution to the torsion flux $\int e^a_k dX^k$:
\begin{equation}
\frac{\beta}{2} b^a \frac{1-\sigma}{2}
\end{equation}
The other contribution corresponding to the discontinuous part of $u$ originates from Eq. (\ref{tsing}). The sum of the two gives
\begin{eqnarray}
\int_{\cal S} T^a_{12} d^2 X \equiv \int_{\partial S} e^a_k dX^k &\approx &  b^a\Big(1-\frac{\beta(1+\sigma)}{4}\Big) \label{torsflux}
\end{eqnarray}
We combine this with Eq. (\ref{ucontsol4}) and obtain:
\begin{eqnarray}
T_{12}^a(X) &=&   b^a \Big( 1-\frac{\beta(1+\sigma)}{4} \Big)\delta^{(2)}(X) + \frac{\beta(1+\sigma)}{2} \frac{\delta^{ik} - 2\hat{X}^i\hat{X}^k}{2\pi |X|^2}b^l {\bf i}^{il} {\bf i}^{ak} \label{ucontsol4}
\end{eqnarray}}

It is worth mentioning, that for the graphene layer on the substrate with the strong connection between the layer and the substrate  $u_{\rm cont}^a$ is given as external condition. If it is regular, the singularity of torsion concentrated on the dislocation is given by Eq. (\ref{tsing}) and instead of Eq. (\ref{ucontsol4}) we have: $T_{12}^a(X) \approx   b^a \Big( 1-\frac{\beta}{2} \Big)\delta^{(2)}(X) + {\rm regular}\, {\rm terms}$.

\subsection{Observation of torsion singularity}

In order to probe torsion singularity the scattering of a quasi - particle on the dislocation may be used. Let us estimate the effect of torsion singularity on the scattering process.  The wave function of the particle satisfies Pauli equation
\begin{equation}
\frac{1}{2} \gamma^a \Big( e \, e_a^{\mu} i D_{\mu} +  i D_{\mu} e\, e_a^{\mu}\Big)\Psi = 0,  \quad \mu,a = 0,1,2,\quad \gamma^0 = \sigma^3, \gamma^1 = - i \sigma^1, \gamma^2 = - i \sigma^2 \label{pauli}
\end{equation}
The covariant derivative $D_{\mu}$ contains emergent magnetic field (the corresponding charges have different signs for the two valleys ${\bf K}^+$ and ${\bf K}^-$).
Recall, that the vielbein is given by:
\begin{eqnarray}
 e_a^{\mu} & = & \delta_a^{\mu} + \partial_a u^{\mu} -\beta \epsilon_{a\mu} + \frac{\beta}{2} \epsilon_{ll} \delta_a^{\mu}, \quad a,\mu = 1,2, \quad e_0^{\mu}  = \frac{1}{v_F}(1+\frac{\beta}{2} \epsilon_{ll})\delta_0^\mu\nonumber\\
 e^a_{\mu} & = & \delta^a_{\mu} - \partial_\mu u^{a} +\beta \epsilon_{a\mu} - \frac{\beta}{2} \epsilon_{ll} \delta^a_{\mu}, \quad a,\mu = 1,2,  \quad e_\mu^{0}  = {v_F}(1-\frac{\beta}{2} \epsilon_{ll})\delta^0_{\mu}\nonumber\\
 e & = & {\rm det} \, e^a_{\mu} = v_F(1-\frac{\beta+2}{2} \epsilon_{ll})
\end{eqnarray}
We are always able to rescale time so that the rescaled $v_F$ is equal to $1$. (We shall imply this rescaling in the present subsection  for simplicity.)
 Let us represent
\begin{equation}
\Psi({\bf r},t)=\Psi_{\bf k}({\bf r},t) e^{-i \int^{({\bf r},t)}_{({\bf r}_0,t_0)} k_a e^a_{\mu}(y) d y^\mu \pm i \int^{({\bf r},t)}_{({\bf r}_0,t_0)} A_{\mu}(y) d y^\mu}
\,
 \label{psi}
\end{equation}
where $k_\mu=(k_0,k_x,k_y)$. The signs denoted by $\pm$ in this expression are opposite for different valleys.  The integral here is along the trajectory ${\cal C}^{({\bf r},t)}_{({\bf r}_0,t_0)}$ given by the function $y^{\mu}(s,x)$. It is parametrized by $s$ and depends on the endpoint $x^{\mu} = ({\bf r},t)$. We assume, that $\Psi_k$ is slow varying, that is its derivatives are much smaller, than the components of $k_a$. {Then vector $k_a$ plays the role of the three - momentum of the incoming particle because around the point $({\bf r}_0,t_0)$ situated far from the dislocation the vielbein $e^a_{\mu} \approx {\rm diag}(v_F, 1,1) $ is flat, and
\begin{equation}
\Psi \sim e^{-i k_0 (t-t_0) v_F + i {\bf k} ({\bf r}-{\bf r}_0)}
\end{equation}}
Here the eigenvalue of the $2$ - momentum $-i \nabla$ is denoted by ${\bf k}_a = - k_a$.  In our case Eq. (\ref{psi}) actually defines a series of different solutions that correspond to the difference in a winding of ${\cal C}$ around the origin of the dislocation. Phase $i \int^{({\bf r},t)}_{({\bf r}_0,t_0)} k_a e^a_{\mu}(y) d y^\mu$ in Eq. (\ref{psi}) may be interpreted in a simple way: the translation between the two points $x^{\mu}_0, x^{\mu}$ in Weitzenbock space is defined as  $R^a=\int^{({\bf r},t)}_{({\bf r}_0,t_0)} e^a_{\mu}(y) d y^\mu$. Plane wave in  this space is determined by phase $k_a R^a$. Substituting Eq. (\ref{psi}) to Eq.(\ref{pauli}) one obtains
\begin{equation}
\Big(\gamma^a k_a +  \gamma^a    e_a^{\mu}\Big[ i \partial_{\mu}  + \int^{({\bf r},t)}_{({\bf r}_0,t_0)} \Big(k_a T^a_{\nu\rho}(y) - F_{\nu\rho}\Big)\frac{\partial y^{\nu}(s,x)}{\partial x^{\mu}} d y^\rho  + i \Gamma_\mu \Big]\Big) \Psi_{\bf k}({\bf r},t)= 0 \,,
\label{pauli}
\end{equation}
We introduce the pseudo - $U(1)$ field $\Gamma_\mu = \frac{1}{2 \, e} e^a_\mu \nabla_{\nu} e\, e_a^{\nu} $. This field may also be interpreted as spin connection.

 We can always choose the trajectories $y(s,x)$ far from the origin of the dislocation (at the distance $|X| \gg a$.) For the vielbein given by Eq. (\ref{elab}) with $u^a$ given by Eq. (\ref{ucontsol1}) we have  $\nabla_{\mu} e\, e_a^{\mu} \sim \frac{1}{|X|^2}$. The non - singular part of torsion is also $\sim \frac{1}{|X|^2}$.  Therefore, the second term in Eq. (\ref{pauli}) may be neglected and we arrive at constant vector $\Psi_k$ that satisfies
  $\gamma^a k_a \Psi_k = 0$. Thus, far from the origin of the dislocation Eq. (\ref{psi}) with constant $\Psi_k$ gives the solution of Pauli equation.

The origin of the dislocation plays the role of a hole in $2D$ surface of graphene (or a line in $2+1$ D space - time) carrying singularity of torsion $T_{12}^a = b^a \Big(1-\frac{\beta(1+\sigma)}{4}\Big) \delta^{(2)}(X)$. The difference between the phases of the two solutions defined by paths ${\cal C}^{(i)}$ and ${\cal C}^{(j)}$ ended at the same point $x^{\mu}$ is defined by the winding number $K_{ij}$ of the contour ${\cal C}^{(ij)} = {\cal C}^{(i)} - {\cal C}^{(j)}$ around the dislocation. The two given solutions differ by the phase factor:
 \begin{equation}
\Psi^{(i)}_k(X) = \Psi^{(j)}_k e^{-i \int_{{\cal C}^{(ij)}} e_{\mu}^a k_a dx^{\mu} \pm  i \int_{{\cal C}^{(ij)}} A_{\mu} dx^{\mu} } =  \Psi^{(j)}_k e^{ i K^{ij} b^a\Big({\cal K}_a  - {\bf k}_a\frac{\beta(1+\sigma)}{4}\Big)  }, \quad {\cal K}_a = \pm {\bf K}^{(0)}_a  + {\bf k}_a \label{psi2}
\end{equation}
Here ${\cal K}$ is the total momentum of the quasiparticle (that is the sum of the unperturbed Fermi point $ \pm {\bf K}^{(0)}$ and ${\bf k}$). For vanishing Gruneisen parameter $\beta = 0$ we are left with especially simple result: for the contour winding once around the origin of the dislocation the phase is equal to $b^a {\cal K}_a$.

The description of the scattering problem of quasiparticle on the dislocation involves these multiple - defined wave functions. We do not give here the detailed description of the quantum - mechanical solution for the given scattering problem.
Let us notice only that this situation is similar to that of the Aharonov - Bohm effect. However, now momentum $k_a$ enters the expression for the phase factor of Eq. (\ref{psi2}). This complicates the consideration. For the elastic scattering with small angles (so that $k_a$ for the incoming and outgoing waves are equal) the analogy is most valuable.
We refer to the appearance of the phase in the wave function proportional to the product of winding number, particle momentum, and torsion concentrated within the loop as to Stodolsky effect (see below).

Measuring of the beam phase shift in the gravitational field (that is expressed  through both gravitational constant and Plank constant) was suggested by Colella and Overhauser in 1974 \cite{Overhauser}. This effect was discussed later in a number of papers (see, for example, \cite{Anandan,stodolsky}). Stodolsky \cite{stodolsky} derived an expression for the phase shift in the case of the gravitational field of general (non - Newtonian) form that reveals an analogy to the Aharonov - Bohm effect. Therefore, we feel it reasonable to call the appearance of this phase (proportional to the particle momentum) the {\it Stodolsky effect}.  Notice, that in the form discussed here the condensed matter analogue of this effect differs from the usual gravimagnetic Aharonov - Bohm effect originated from the components $g_{0k}$, $k \ne 0$ \cite{Volovik2003}. The observation of this effect in graphene for the scattering of quasi-particle on the torsion singularity carried by the dislocation was proposed in \cite{Guinea2014} (in this respect in \cite{Guinea2014} the first term of Eq. (\ref{using}) was discussed while the contribution of $u_{\rm cont}$ was not considered).

\section{Discussion}

As distinct from the fundamental gravity, in the effective elastic gravity there is no diffeomorphism invariance. The coordinate transformation describes real deformation of the lattice, which modifies the distances between the atoms. In the so - called "laboratory reference frame" (in which the points of the surface are parametrized by the projections of their coordinates to a "laboratory" plain) there are two sources of the response of the fermionic gravity to the elastic gravity: the geometric one, which is determined by the coordinate transformation; and the material one, which is determined by the material  Gr\"uneisen parameters. At the same time in the "accompanying reference frame" (in which the coordinates of the atoms are equal to their coordinates in the unperturbed lattice) there is only the material contribution: the "geometric" one is eliminated by the transformation between the two coordinate systems.

This should be taken into account, when the geometric methods are applied to the study of properties of topological solid-state or liquid materials. There the fictitious gravitational field is introduced in the Newton-Cartan formulation
with localization of Galilean invariance, see e.g. recent papers \cite{BradlynRead2014,MorozHoyos2014,Jensen2014,GromovAbanov2014,Andreev2014}.

\revision{Our consideration of emergent geometry in the presence of dislocations demonstrates that
the dislocation carries the singularity of torsion in addition to the finite flux of emergent magnetic field. The latter may be observed via the Aharonov - Bohm effect. The torsion singularity may also be probed by the scattering of quasipartcles on the dislocation. It manifests itself in Stodolsky effect that is similar to the Aharonov - Bohm effect, i.e. the quasipaticle winding around the torsion singularity acquires the phase that is proportional to the torsion localized on the dislocation multiplied by the momentum of the incoming wave.
The observation of a similar phase shift in general relativity was suggested in \cite{Overhauser} and discussed in a number of papers (see, for example, \cite{Anandan,stodolsky}). Stodolsky \cite{stodolsky} discussed the non - Newtonian case, when the expression for the phase shift may be easily generalized to take the form considered here. In \cite{Guinea2014} it was proposed to observe this effect in graphene for the scattering of quasi - particles on the dislocation. However, in \cite{Guinea2014} only the torsion originated from the discontinuous part of the displacement vector (the first term of Eq. (\ref{using})) was discussed. In the present paper we take into account torsion singularity originated from the second term of Eq. (\ref{using}) given by the elasticity equations. This results in a correction to the phase acquired by the particle proportional to material parameter $\beta$. Notice, that unlike for the usual Aharonov - Bohm effect, the phase acquired by the particle when it surrounds torsion singularity is proportional to its momentum and depends on the direction of momentum.

The particular distribution of dislocations simulates effective emergent magnetic field and effective torsion distributed within the graphene plane (see also \cite{Dzyaloshinskii1980}). In that case in addition to the torsion field originated from the derivative of the effective vielbein $e^a_k$ the extra contribution appears that is related to the distribution of dislocations. Such a contribution may be interpreted through the appearance of spin connection $\Big[\omega_i\Big]^a_b$ defined through relation
\begin{equation}
T^a_{ij} = \partial_{[i}E^a_{j]} + T^a_{ij, {\rm dislocations}} \equiv \Big[D_{[i}E_{j]}\Big]^a, \quad D_i= \partial_i  + \omega_i
\end{equation}
(Here $E^a_i$ is the regular part of inverse vielbein. It is equal to $e^a_i$ with the singular contribution of dislocations subtracted.)}

The interrelation between the two types of  metric considered here on an example of graphene may be extended to the other condensed matter systems with Fermi points in the presence of elastic deformations. The example of such a $2D$ system is given by the boundary of topological insulator. The example of the $3D$ system is given by the Weyl and Dirac semimetals \cite{Abrikosov1971,Abrikosov1998,Burkov2011,XiangangWan2011,Weylsemimetal}. In both these cases there exists the natural metric of elasticity theory that is determined by strain. However, the fermionic quasi - particles experience another metric given by the expansion of the effective Hamiltonian near Fermi point. The particular expressions for the emergent vielbein and emergent gauge field depend on the details of the considered systems and differ from the expressions for graphene considered above. Notice, that unlike graphene in these cases the emergent spin connection may appear in addition to the emergent gauge field (in the absence of dislocations).

The work of M.A.Z. is  supported by the Natural Sciences and Engineering
Research Council of
Canada and by grant RFBR 14-02-01261. GEV
acknowledges a financial support of the Academy of Finland and its COE
program.

 \end{document}